\documentclass[toc]{PoS}
\usepackage{graphicx}

\title{Numerical Evaluation of Feynman Integrals by a Direct Computation Method}

\ShortTitle{Numerical Evaluation of Feynman Integrals by a Direct Computation Method}

\author{\speaker{F. Yuasa}$^a$, T. Ishikawa$^a$, J. Fujimoto$^a$,
N.Hamaguchi$^b$, E. de Doncker$^c$, Y. Shimizu$^d$\\
\llap{$^a$}High Energy Accelerator Research Organization (KEK), 1-1 OHO Tsukuba, Ibaraki  305-0801, Japan\\
\llap{$^b$}Hitachi, Ltd., Software Division, Totsuka-ku, Yokohama, 244-0801, Japan\\
\llap{$^c$}Western Michigan University, Kalamazoo, MI 49008-5371, USA\\
\llap{$^d$}The Graduate University for Advanced Studies, Sokendai, Shonan Village, Hayama, Kanagawa 240-0193, Japan\\
E-mail: \email{fukuko.yuasa@kek.jp},\\
\email{tadashi.ishikawa@kek.jp},\\
\email{junpei.fujimoto@kek.jp}, \\
\email{nobuyuki.hamaguchi.sa@hitachi.com},\\
\email{elise@cs.wmich.edu},\\
\email{shimiz@suchix.kek.jp}}
       
\abstract
{A purely numerical method, {\it Direct Computation Method} is applied to evaluate 
Feynman integrals. This method is based on the combination of an efficient 
numerical integration and an efficient extrapolation. In addition, high-precision 
arithmetic and parallelization technique can be used in this method if required.
We present the recent progress in development of this method and show results 
such as one-loop 5-point and two-loop 3-point integrals.
}

\FullConference{ACAT2008 \\
XII International Workshop on Advanced Computing and Analysis Techniques in Physics Research\\
Erice, Sicily, Italy\\
November 3-7, 2008}

\begin{document}
\twocolumn
%
\section{Introduction}\label{sec:intro}
This paper describes a computational method for the loop integrals. 
This method is named as {\it Direct Computation Method} and its
advantage  is handling singularities in a purely numerical way 
which appear in the denominator of the integrand .
We have already computed several diagrams with one-loop and two-loop such as
\begin{itemize}
\item{One-loop vertex and box diagrams with/\\without infrared divergence\cite{dq1,dq2,dq3,dq4}},
\item{Two-loop planar and non-planar vertex \\diagram\cite{dq5}}.
\end{itemize}
In this paper, we present the brief description of this method and show 
another examples of loop integrals.
\section{Direct Computation Method}\label{sec:directcomputation}
This method is based on a numerical multi-dimensional integration to get 
the sequence of the integration approximations of the loop integral, $\{I(\epsilon_{l})\},l=0,1,2,...$, 
and an extrapolation technique for the convergence of the sequence.
Here, the sequence of $\{\epsilon_{l}\},l=0,1,2,...$  is given as $\epsilon_{l} = \epsilon_{0}\times a^{-l}$ and $\epsilon_{0}$ and $a$ are real constants such as 256 and 2 respectively\cite{dq1}. 
When we take the limit of $\epsilon \rightarrow 0$, we get the result of the loop integral.

In {\it Direct Computation Method}, the integration routine plays a central role.
We have been using two different integration routines, {\tt DQAGE} and {\tt DE}.
The former, {\tt DQAGE}, is an adaptive algorithm routine in {\tt QUADPACK} \cite{quadpack}.
We call the combination of {\tt DQAGE} and the extrapolation technique {\it DQ-Direct Computation Method} in this paper. 
The latter, {\it Double Exponential formulas}\cite{de}, shortly {\tt DE}, uses $tanh(\pi/2*sinh(t))$ transformation for numerical integration. We call the combination of {\tt DE} and the extrapolation technique {\it DE-Direct Computation Method} in this paper. 

For both integration routines, we are combining Wynn's $\epsilon$ algorithm \cite{dq1,wynn} or Aitken extrapolation to accelerate the convergence. 
\section{Examples of the Computation}
\subsection{One-loop box contributing to $gg \rightarrow b {\bar b} H$}
\begin{figure}[h]
\caption{Box diagram contributing to $gg \rightarrow b {\bar b} H$}
\label{fig:bbh}
\begin{center}
\includegraphics[width=5.0cm]{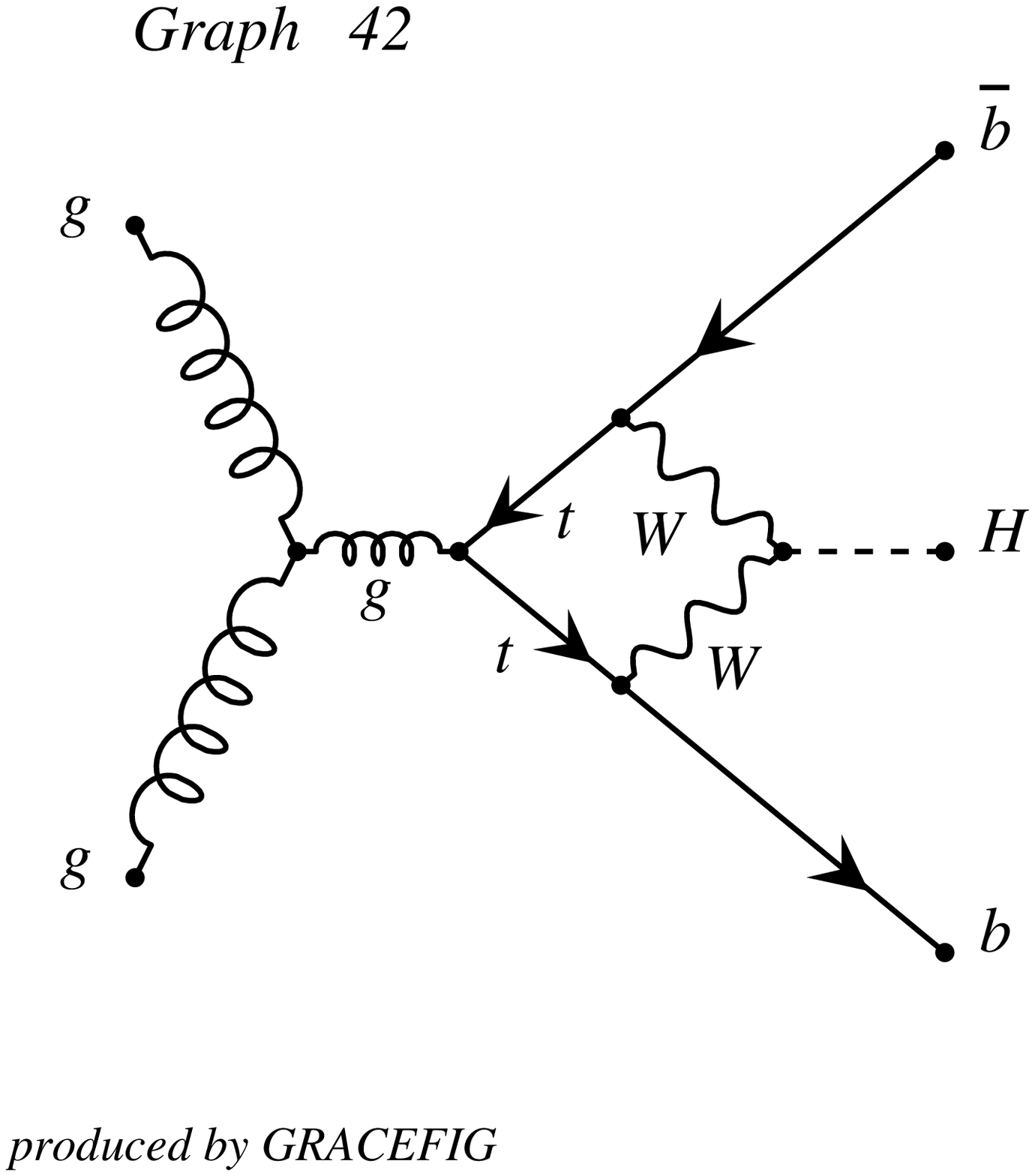}
\end{center}
\end{figure}

\begin{figure}[h]
\caption{Integration result of the real part with complex masses as a function of $\sqrt{s_{2}}$. Our results agree to the 5- or 6-digit accuracy of Ninh's.}
\label{fig:bbhreal}
\begin{center}
\includegraphics[width=6.0cm]{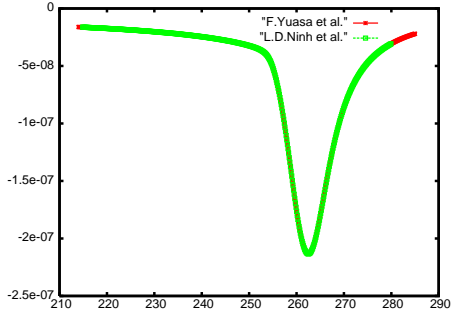}
\end{center}
\end{figure}

\begin{figure}[h]
\caption{Integration result of the imaginary part with complex masses as a function of $\sqrt{s_{2}}$. Our results agree to the 4-, 5- or 6-digit accuracy of Ninh's.}
\label{fig:bbhimag}
\begin{center}
\includegraphics[width=6.0cm]{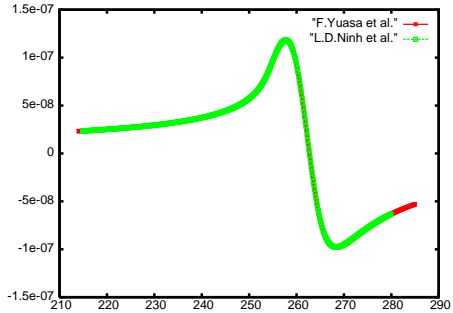}
\end{center}
\end{figure}

The first example is the the one-loop box diagram with complex masses contributing to $gg \rightarrow b {\bar b} H$ in Fig.~\ref{fig:bbh}.
\vspace{-0.1cm}
In the computation we take$\sqrt{s}=353$ GeV as an example and we fix $\sqrt{s_{1}}=\sqrt{2(m_t^2+M_W^2)}\approx271.06$ GeV. As for the mass parameters, $m_{t}=174$ GeV, $M_{W}=80.3766$ GeV and $M_{H}=165$ GeV.
We introduce the complex masses as $m_{t}^{2} \rightarrow m_{t}^{2}-im_{t}\Gamma_{t}$ and $M_{W}^{2} \rightarrow M_{W}^{2}-iM_{W}\Gamma_{W}$ with $\Gamma_{t} = 1.5$ GeV and $\Gamma_{W} = 2.1$ GeV. The results of the numerical computation agree perfectly to the analytic results reported by L. D. Ninh {\it et al.} \cite{bbh} as in Fig.~\ref{fig:bbhreal} and Fig.~\ref{fig:bbhimag}. 

\subsection{$e^{+}e^{-} \rightarrow ZZ \rightarrow e^{+}e^{-}Z$}
\begin{figure}[h]
\caption{Diagram of $e^{+}e^{-} \rightarrow ZZ \rightarrow e^{+}e^{-}Z$}
\label{fig:1lp}
\begin{center}
\includegraphics[width=5.0cm]{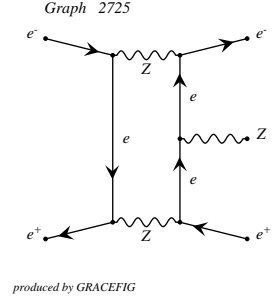}
\end{center}
\end{figure}

The second example is the one-loop pentagon diagram in
Fig.~\ref{fig:1lp}. 
%
%
The loop integral is
\begin{equation}
I = \int_{0}^{1} dx_{1}dx_{2}dx_{3}dx_{4}dx_{5}\delta(1-\sum_{i=1}^{5}x_{i})\frac{1}{D^{3}}.
\end{equation}
Here, $D({\bf x})$ is given as
\begin{eqnarray}
\label{eqn:pentagon}
D({\bf x}) &=& \sum_{i=1}^{5} m_{i}^{2}\nonumber\\
&-& x_{1}x_{2}s_{1} - x_{2}x_{3}s_{2} - x_{3}x_{4}s_{3} - x_{4}x_{5}s_{4}\nonumber\\
&-& x_{5}x_{1}s_{5} -x_{1}x_{3}s_{12} - x_{2}x_{4}s_{23} -x_{3}x_{5}s_{34}\nonumber\\
&-&x_{4}x_{1}s_{45} - x_{5}x_{2}s_{51},
\end{eqnarray}
where $m_{1}=m_{3}=m_{4}=0.5\times10^{-3}$ GeV and $m_{2}=m_{5}=91.00$ GeV.
As an example, one numerical result is shown in Table~\ref{tab:pentagon} with a set of parameters corresponding to one phase space point (Table~\ref{tab:1lp}).
\begin{table} 
\caption{Numerical result of one-loop pentagon.}
\label{tab:pentagon}
\begin{center}
\begin{tabular}{lcc}\hline
& Result & Error \\ \hline
real & $ 0.4118519 \times 10^{-13}$ & $0.410 \times 10^{-17}$ \\
imag.& $-0.2336871 \times 10^{-12}$ & $0.619 \times 10^{-16}$ \\ \hline
\end{tabular}
\end{center}
\end{table}
These agree to the results by T. Ueda {\it et al.} \cite{ueda-acat08}.
\begin{table}[h]
\caption{A set of parameters corresponding to one phase space point}
\label{tab:1lp}
\begin{center}
\begin{tabular}{l|l}\hline
Parameter& value \\ \hline
$S_{12}$&100000.00000\\
$S_{15}$&-14146.0960752976\\
$S_{23}$&-30471.3126018059\\
$S_{34}$&32384.1496580698\\
$S_{45}$&37833.5682283554\\ \hline
\end{tabular}
\end{center}
\end{table}
The elapsed time for the computation is about 9.3 hours for the real part using AMD Opteron 2.2 GHz CPU. 
\subsection{Two-Loop Self-energy}
\begin{figure}[h]
\caption{Diagram of the two-loop self-energy}
\label{fig:2ls}
\begin{center}
\includegraphics[width=4.5cm]{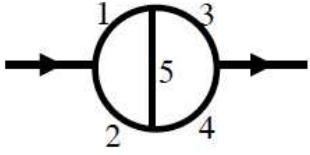}
\end{center}
\end{figure}

The third example is the two-loop self-energy diagram in Fig.~\ref{fig:2ls}. 

The formulae of the loop integral is \cite{2loopself}
\begin{equation}
(16\pi^{2})^{2}I = \int_{0}^{1} dx_{1}dx_{2}dx_{3}dx_{4}dx_{5}\delta(1-\sum_{i=1}^{5}x_{i})\frac{1}{CD}.
\end{equation}
Here, $D({\bf x})$ is given as
\begin{eqnarray}
D({\bf x})&=& -p^{2}(x_{5}(x_{1}+x_{3})(x_{2}+x_{4})\nonumber\\
&+&(x_{1}+x_{2})x_{3}x_{4}+(x_{3}+x_{4})x_{1}x_{2})\nonumber\\
&+&C{\tilde M^{2}},
\end{eqnarray}
\begin{eqnarray}
C&=&(x_{1}+x_{2}+x_{3}+x_{4})x_{5}+(x_{1}+x_{2})(x_{3}+x_{4}),\nonumber\\
{\tilde M^{2}} &=& \sum_{i=1}^{5} x_{i}m_{i}^{2}.
\end{eqnarray}

\begin{figure}[h]
\caption{Numerical results of the loop integral with Set\#1 mass assignment as a function of $p^{2}/150.0^{2}$. The results of \cite{kreimer,nci} are divided by the value of $p^2$.}
\label{fig:self-case1}
\begin{center}
\includegraphics[width=6.0cm]{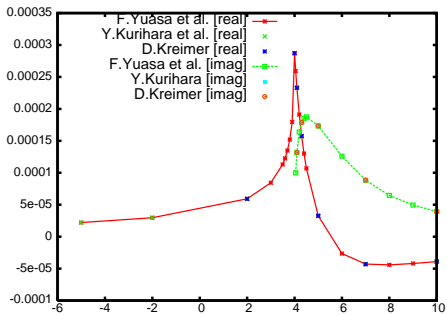}
\end{center}
\end{figure}

\begin{figure}[h]
\caption{Numerical results of the loop integral with Set\#2 mass assignment as a function of $p^{2}$. The sign of our numerical results is conformed to \cite{nci,bauberger,passarino}. }
\label{fig:self-case2}
\begin{center}
\includegraphics[width=6.0cm]{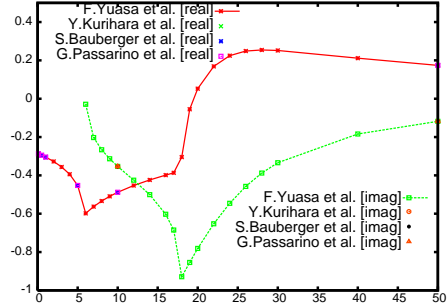}
\end{center}
\end{figure}

We show the numerical results with two sets of the mass assignment shown in 
Table~\ref{tab:2ls-param}. The results with the first set are compared 
with ones by Kreimer \cite{kreimer} and by Kurihara {\it et al.} 
\cite{nci} in Fig.~\ref{fig:self-case1}.
The results with the second set are compared with ones by Kurihara {\it et al.} 
\cite{nci}, by Bauberger {\it et 
al.} \cite{bauberger} and by Passarino {\it et al.} \cite{passarino} in Fig.~\ref{fig:self-case2}. 
In this computation, the elapsed time for the computation becomes longer around the singularities.

\begin{table}[h]
\caption{Set of mass assignment in GeV}
\label{tab:2ls-param}
\begin{center}
\begin{tabular}{clllll}\hline
Set\#&$m_{1}$&$m_{2}$&$m_{3}$&$m_{4}$&$m_{5}$\\ \hline
1   &150.0&150.0&150.0&150.0& 91.17\\
2   &$\sqrt 1.0$&$\sqrt 2.0$&$\sqrt 4.0$&$\sqrt 5.0$& $\sqrt 3.0$\\ \hline
\end{tabular}
\end{center}
\end{table}
\vspace{-0.5cm}

%
\normalsize
\section{Parallel Computation}
\begin{figure}
\caption{Speedup of the parallel code of {\it DE-Direct Computation Method} for the one-loop box integral with ({\tt DE-xd1-s-positive}) and without ({\tt DE-xd1-s-negative}) singularities. We use CRAY XD1 for the measurement.}
\label{fig:DE-xd1}
\begin{center}
\includegraphics[width=6.0cm]{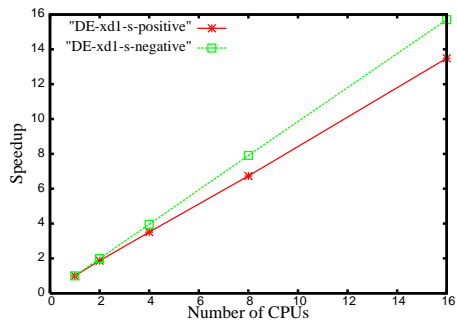}
\end{center}
\end{figure}

\begin{figure}
  \caption{Speedup of the parallel code of {\it DE-Direct Computation Method} for the one-loop box integral with ({\tt DE-bg-s-positive}) and without ({\tt DE-bg-s-negative}) singularities. We use IBM BlueGene/L at KEK for the measurement.}
\label{fig:DE-bg}
\begin{center}
\includegraphics[width=6.0cm]{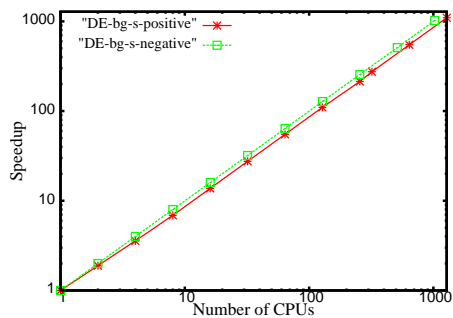}
\end{center}
\end{figure}

As we have described, the elapsed time for the computation becomes longer to get 
the sequence $\{I(\epsilon_{l})\}$ in a good accuracy when the singularity
becomes steeper. To reduce the computation time, parallel computing technique is often used.
We have developed the parallel code of {\it Direct Computation Method.}
We used {\tt MPI}\footnote{Message Passing Interface} library which is widely used in the  parallel computing environment. 
We evaluated the speedup of the parallel code of {\it DQ -} and {\it DE-} {\it Computation Method} taking the integrals of two-loop vertex and one-loop box diagram as examples. 
The results of the speedup are shown in Table~\ref{tab:DQ-parallel}, Fig.~\ref{fig:DE-xd1} and Fig.~\ref{fig:DE-bg}. Both parallel codes show a good speedup behavior. 

\begin{table}
\caption{Speedup of the parallel code of {\it DQ-Direct Computation Method} for the two-loop vertex diagram with singularities.
We use PPC 970 and PC farm for the measurement. PPC 970 is IBM Blade CenterJS20 with PowerPC 970 2.2 GHz CPUs. PC farm consists of 16 Intel Xeon 3.06 GHz CPUs. }
\label{tab:DQ-parallel}
\begin{center}
\begin{tabular}{l|l|l}\hline
\# of CPUs & PPC 970& PC farm \\ \hline
1&1.00&1.00 \\ 
2&1.53&1.50 \\ 
4&2.99&2.54 \\ 
8&5.50&4.54 \\ 
16&&13.31 \\ \hline 
\end{tabular}
\end{center}
\end{table}

\section{Summary}\label{sec:summary}
In this paper, we presented several numerical results of the loop integrals by  
{\it Direct Computation Method}. This method is based on the combination of the numerical integration and an extrapolation technique. 
To reduce the computation time we developed the parallel code and 
the results of the speedup measurement are also shown.
\acknowledgments
We wish to thank Dr. Kurihara and Prof. Kaneko
for their valuable suggestions. We wish to thank Prof. Kawabata for his support. 
This work is supported in part by the project of Hayama Center
for Advanced Studies at the Graduate University for Advanced
Studies.

\end{document}